\documentclass[manuscript]{aastex}

\slugcomment{accepted for publication by The Astrophysical Journal}

\shorttitle{Giant-Impact Induced Debris Disks}
\shortauthors{Genda et al.}

\begin{document}


\title{Warm Debris Disks Produced by Giant Impacts During Terrestrial Planet Formation}


\author{H. Genda\altaffilmark{}}
\affil{Earth-Life Science Institute, Tokyo Institute of Technology, 
       Ookayama, Meguro-ku, Tokyo, 152-8550, Japan}
\email{genda@elsi.jp}

\author{H. Kobayashi\altaffilmark{}}
\affil{Department of Physics, Graduate School of Science, Nagoya University,
       Furo-cho, Chikusa-ku, Nagoya, 464-8602, Japan}

\and

\author{E. Kokubo\altaffilmark{}}
\affil{Division of Theoretical Astronomy, National Astronomical Observatory of Japan, 
       Osawa, Mitaka, Tokyo 181-8588, Japan}

\begin{abstract}

In our solar system, Mars-sized protoplanets frequently collided with each other during the last stage of terrestrial planet formation called the giant impact stage. Giant impacts eject a large amount of material from the colliding protoplanets into the terrestrial planet region, which may form debris disks with observable infrared excesses. Indeed, tens of warm debris disks around young solar-type stars have been observed. Here, we quantitatively estimate the total mass of ejected materials during the giant impact stages. We found that $\sim$0.4 times the Earth's mass is ejected in total throughout the giant impact stage. Ejected materials are ground down by collisional cascade until micron-sized grains are blown out by radiation pressure. The depletion timescale of these ejected materials is determined primarily by the mass of the largest body among them. We conducted high-resolution simulations of giant impacts to accurately obtain the mass of the largest ejected body. We then calculated the evolution of the debris disks produced by a series of giant impacts and depleted by collisional cascades to obtain the infrared excess evolution of the debris disks. We found that the infrared excess is almost always higher than the stellar infrared flux throughout the giant impact stage ($\sim$100 Myr) and is sometimes $\sim$10 times higher immediately after a giant impact. Therefore, giant impact stages would explain the infrared excess from most observed warm debris disks. The observed fraction of stars with warm debris disks indicates that the formation probability of our solar system-like terrestrial planets is approximately 10\%.

\end{abstract}





\section{Introduction}

Collisions among Mars-sized protoplanets have been thought to be common events during terrestrial planet formation in our solar system \citep[e.g.,][]{Chambers1998, Agnor1999}, and these collisions are called giant impacts. In the solar system, several Mars-sized protoplanets formed in the terrestrial planet region ($\sim$ 1AU from the Sun) through the successive accretion of planetesimals \citep[e.g.,][]{Wetherill1985, Kokubo1998, Kobayashi2010b, Kobayashi2013}. Giant impacts among protoplanets are triggered by the depletion of the gas in the protoplanetary disk, which leads to the instability of the orbits of the protoplanets. The stage in which many giant impacts occur is known as the giant impact stage and lasts approximately 100 Myr in the solar system \citep[e.g.,][]{Jacobson2014, Chambers1998}. This is consistent with the timing of the formation of the Moon, as inferred from its geochemistry \citep{Touboul2007, Halliday2008}.

Even in extrasolar planetary systems, giant impact events are thought to be common. For example, short-period super-Earths have been thought to grow through collisions among Mars- and Earth-sized protoplanets that migrate inward to the vicinity of the central stars \citep{Terquem2007}. These giant impacts in extrasolar planetary systems after the depletion of the gas in the protoplanetary disk can produce the observed architecture of super-Earth systems where short-period super-Earths are not in mean-motion resonance \citep{Ogihara2009}. \cite{Ogihara2013, Ogihara2014} also investigated the possibility that super-Earths grow through giant impacts outside a hot Jupiter. They found that growing super-Earths push the hot Jupiter inward, and eventually the hot Jupiter is swallowed into the central star, which may explain the lack of super-Earths in systems with a hot Jupiter.

Giant impacts significantly influence various features of planetary systems, such as the number, mass, and spin of terrestrial planets \citep[e.g.,][]{Agnor1999, Kokubo2006}. Giant impacts are highly energetic events and are responsible for the creation of the Moon \citep[e.g.,][]{Canup2004, Genda2003a}, the martian satellites \citep{Citron2015}, and planets with extremely large cores such as Mercury \citep[e.g.,][]{Benz2007}. Moreover, giant impacts are closely related to the thermal states of terrestrial planets, such as magma oceans \citep[e.g.,][]{Tonks1992, Hamano2013}, and the origins of terrestrial planet atmospheres \citep{Genda2003b, Genda2005}.

According to recent simulations of giant impacts, there are several types of collision outcomes, including perfect merging, hit-and-run, partial accretion, and disruptive collisions \citep{Agnor2004, Leinhardt2012, Genda2012, Sekine2012}. Even in the case of a low-velocity giant impact ($v_{\rm imp} \approx v_{\rm esc}$, where $v_{\rm imp}$ and $v_{\rm esc}$ are the impact velocity and the two-body surface escape velocity, respectively), such as the Moon-forming giant impact, approximately 1\% of the Earth's mass is ejected and escapes from the Earth's gravity \citep[e.g.,][]{Canup2004}. Moreover, if Mercury's large core is the result of giant impact(s), a significant amount of mantle material is ejected from the proto-Mercury \citep{Benz1988, Benz2007, Asphaug2014}. Therefore, it is expected that a large amount of material is ejected into the terrestrial planet region during the giant impact stage \citep[e.g.,][]{Leinhardt2012, Stewart2012}. In this paper, we call the material ejected by a giant impact ``fragments'' regardless of their sizes and masses.

Recently, thanks to infrared space telescopes such as {\it Spitzer} and {\it AKARI}, tens of warm debris disks around solar-type (FGK) stars with ages of $10^7-10^8$ years have been reported \citep[e.g.,][]{Zuckerman2011, Fujiwara2013}. The temperatures of these debris disks have been estimated to exceed 150 K from the spectral energy distribution including infrared excess. These warm debris disks are estimated to be located roughly 1 AU to several AU from the central stars, which corresponds to the terrestrial planet region in the solar system and the super-Earth-forming region around some extrasolar stars. Based on their stellar ages and locations of the debris disks, the relation between these warm debris disks and giant impact events has recently been discussed \citep[e.g.,][]{Weinberger2011, Jackson2012, Melis2012}.

\cite{Jackson2012} considered the Moon-forming giant impact. This impact is an oblique collision between Earth- and Mars-sized protoplanets with a low impact velocity ($\approx v _{\rm esc}$), which led to the formation of a proto-lunar disk, from which the Moon was made, around the Earth-sized protoplanet. They focused on the fragments produced by the Moon-forming giant impact, and performed numerical simulations of the orbital and collisional evolution of the ejected fragments around the Sun. They adopted 1.6\% of $M_{\oplus}$ (the mass of the Earth) as the total mass of the ejected fragments, which was taken from the numerical result of the giant impact simulation \citep{Marcus2009}, and 0.2\% of $M_{\oplus}$ as the largest fragment, which was taken from the largest clump in the proto-lunar disk produced by the Moon-forming giant impact (Canup 2008). The lifetime of a debris disk strongly depends on the mass of the largest fragment. Although the mass of the largest clump in the proto-lunar disk is not exactly the same as that of the largest fragment that is ejected from the potential well of the Earth-sized protoplanet, they estimated that the debris disk produced by the single Moon-forming giant impact has a lifetime of $\sim$ 30 Myr. They concluded that this debris disk can be observed by infrared excess at 24 $\mu$m. 

Many giant impacts occur during the giant impact stage, and there are several types of collision outcomes other than Moon formation. To determine the infrared excess from a debris disk throughout the giant impact stage, all types of giant impacts that take place in the giant impact stage should be taken into account. Moreover, because the total mass of the fragments and the mass of the largest fragment ejected by each giant impact are the important parameters in determining the lifetime of the debris disk, these parameters should be determined via high-resolution giant impact simulations.

In this paper, we calculate the infrared excess from a debris disk produced by multiple giant impacts in the giant impact stage. For this purpose, we need to determine the number and timing of giant impacts, the outcomes of each giant impact, and the collisional evolution of the fragments. The giant impact events in the giant impact stage occur because of the chaotic orbital instability of protoplanets, which must be investigated by {\it N}-body simulations taking the mutual gravity into account. Using the results of {\it N}-body simulations of the giant impact stage by \cite{Kokubo2010} and the smoothed particle hydrodynamics (SPH) simulations of giant impacts by Genda et al. (2012), we calculate the total ejected fragment mass during the giant impact stage, as described in Section \ref{sec:Mass}. The fragments collide with each other, which leads to produce smaller fragments. This collisional cascade grinds down fragments to micron-sized grains, which are blown out by radiation pressure. The depletion timescale of fragments depends on the mass of the largest fragment \citep{Kobayashi2010a}. However, the SPH simulations by \cite{Genda2012} were not performed at a sufficiently high resolution to accurately obtain the mass of the largest ejected fragment. In Section \ref{sec:GI}, we newly conduct high-resolution SPH simulations of giant impacts. In Section \ref{sec:Lifetime}, we investigate the collisional evolution of the fragments using the results obtained in Section \ref{sec:GI}, and we calculate the mass evolution of debris disks. Using these results, we show the infrared flux evolution during the giant impact stage in Section \ref{sec:Excess}. Finally, we discuss our findings in Section \ref{sec:Discussion}.

\section{Total Mass Ejected by Giant Impacts during Giant Impact Stage}\label{sec:Mass}

In this section, we quantitatively estimate the total mass of fragments ejected by giant impacts that occur during the giant impact stage by using the results obtained by \cite{Kokubo2010} and \cite{Genda2012}.

\cite{Kokubo2010} performed {\it N}-body orbital calculations of protoplanets in the giant impact stage up to 200 Myr, taking into account both hit-and-run and merging collisions. For the initial conditions, 16 Mars-sized protoplanets with a total mass of 2.3$M_{\oplus}$ are located from 0.5 to 1.5AU. The orbital separations of protoplanets are given by 10 mutual Hill radii in accordance with the results of planetesimal accretion \citep{Kokubo1998}. The initial eccentricities and inclinations of protoplanets are set to have a Rayleigh distribution with dispersions of 0.01 and 0.005 rad, respectively. In \cite{Kokubo2010}, 50 runs with different initial angular distributions of protoplanets were performed. As the results of statistics, 1211 giant impacts occur in 50 runs. This corresponds to an average of 24 giant impacts per run. The number of finally formed terrestrial planets is $3.6 \pm 0.8$, and the final giant impact occurs at $73 \pm 74$ Myr. These numerical simulations provide us the time, location, and impact parameters (the masses of the two colliding protoplanets, impact velocity, and impact angle) of all giant impacts that take place during the 50 runs of the giant impact stage.

\begin{figure}
\epsscale{1.0}
\plotone{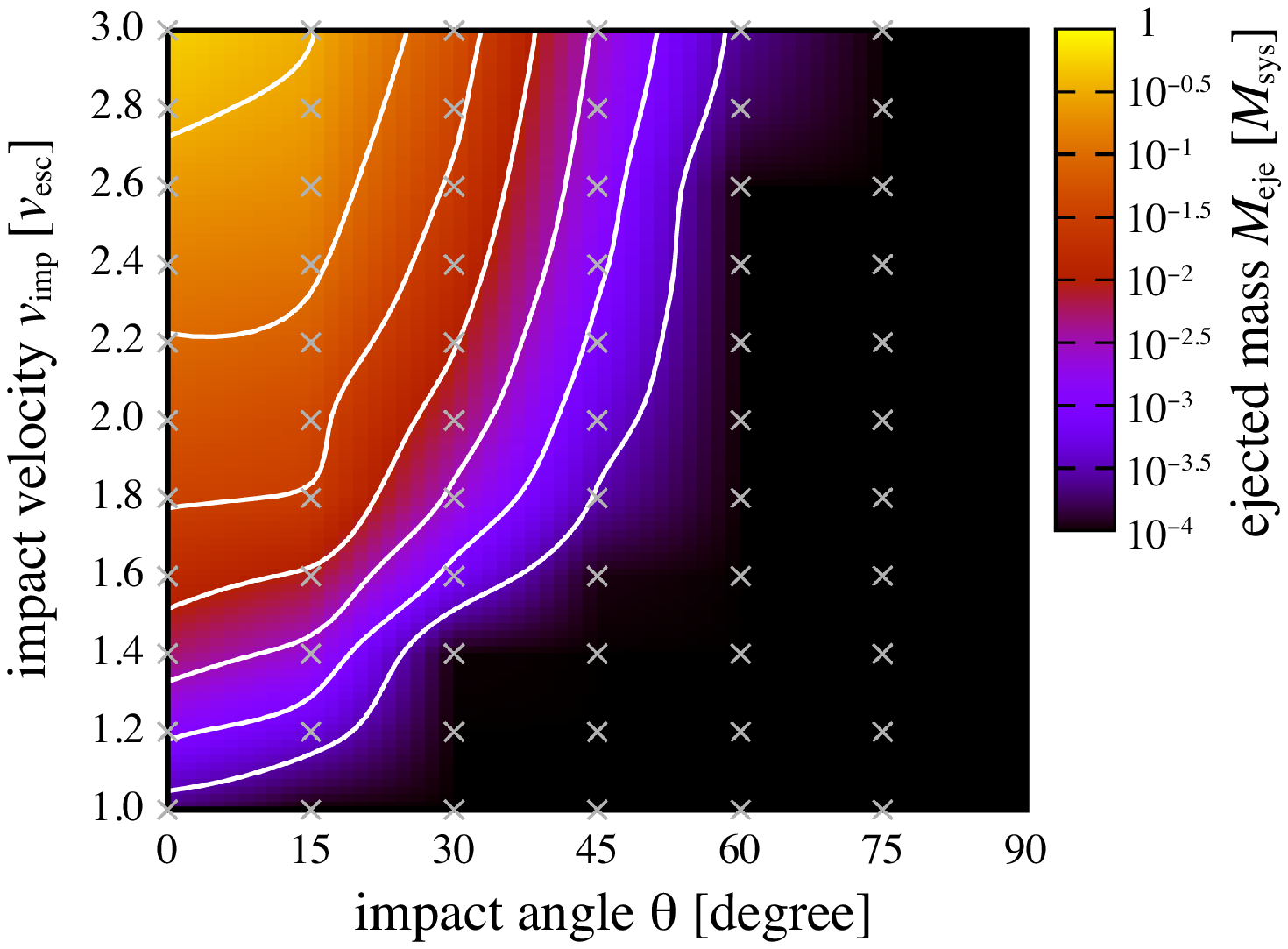}
\caption{The total mass of the ejected fragments $M_{\rm eje}$ by a single giant impact between protoplanets of the same size with a mass of 0.1$M_{\oplus}$ as a function of impact velocity $v_{\rm imp}$ and angle $\theta$. The color contour represents $M_{\rm eje}$ normalized by the total mass of the colliding protoplanets $M_{\rm sys}$. From bottom to top white contours represent $M_{\rm eje}/M_{\rm sys} = 10^{-3.5} - 10^{-0.5}$ with an interval of $10^{0.5}$. The impact velocity is normalized by the two-body surface escape velocity $v_{\rm esc}$. A head-on collision corresponds to $\theta$ = 0. Crosses represent the impact conditions of SPH calculations performed by \cite{Genda2012}.}
\label{fig:Meje}
\end{figure}

\cite{Genda2012} performed SPH simulations of giant impacts between protoplanets with a 30wt\% iron core and a 70wt\% silicate mantle. More than 1000 impact simulations for various impact parameters (the mass ratio of the colliding protoplanets, impact velocity, and impact angle) were systematically carried out, and the merging criteria of protoplanets were investigated. Although \cite{Genda2012} did not focus on the ejected fragments, in this paper, we calculate the total mass of the fragments based on the data obtained by \cite{Genda2012}. Figure \ref{fig:Meje} shows an example of the total mass of the fragments ejected by a single giant impact between protoplanets of the same size with mass 0.1$M_{\oplus}$ as a function of impact velocity, $v_{\rm imp}$, and angle, $\theta$. Collisions with lower angles (near-head-on collisions) and higher impact velocities tend to produce a larger amount of fragments. We only count the ejected fragments that are not gravitationally bounded to the colliding protoplanet(s), and thus the fragments orbiting around the protoplanets, such as a proto-lunar disk, are not included in this mass. We tabulate the data of the ejected mass $M_{\rm eje}$ for various impact parameters.

\begin{figure}
\epsscale{1.0}
\plotone{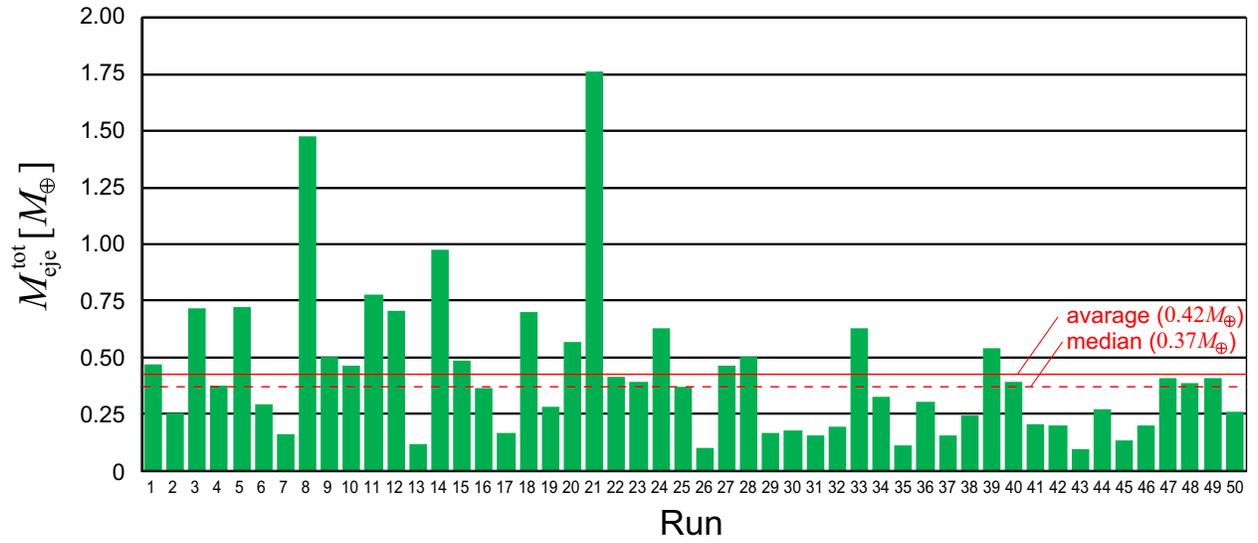}
\caption{Total ejected mass $M_{\rm eje}^{\rm tot}$ normalized by the Earth's mass $M_{\oplus}$ during the giant impact stage for each run.}
\label{fig:Mejetot}
\end{figure}

Using the impact parameters obtained by Kokubo and Genda (2010) and the table data of $M_{\rm eje}$ made here from the results obtained by Genda et al. (2012), we can calculate $M_{\rm eje}$ for all giant impacts that take place in the 50 runs of the giant impact stage. Figure \ref{fig:Mejetot} shows the total mass of the ejected fragments, $M_{\rm eje}^{\rm tot}$, in each run. The average total mass over the 50 runs is $M_{\rm eje}^{\rm tot} = 0.42M_{\oplus}$, which corresponds to 18\% of the total mass of the system ($2.3M_{\oplus}$) and is consistent with the previously estimated value of $15\%$ by \cite{Stewart2012}. The maximum and minimum total masses are $M_{\rm eje}^{\rm tot} = 1.76M_{\oplus}$ and $0.09M_{\oplus}$ and occur during Runs 21 and 26, respectively. On average, each giant impact produces fragments with a total mass of $0.02 M_{\oplus}$. Although the ejected mass by a single giant impact is much smaller than the mass of the terrestrial planets, $M_{\rm eje}^{\rm tot}$ is comparable to or greater than the typical mass of warm debris disks \citep[e.g.,][]{Fujiwara2012}. It should be noted that the ejected mass is reduced by collisional evolution in the debris disk. We investigate the evolution of the debris disk in Section \ref{sec:Lifetime}. 

The values of $M_{\rm eje}^{\rm tot}$ estimated here is somewhat overestimated, because \cite{Kokubo2010} does not take into account the mass loss by giant impacts in their {\it N}-body calculations. We used the masses of perfectly merged protoplanets to estimate each $M_{\rm eje}$. Especially, for the cases with high $M_{\rm eje}^{\rm tot}$ such as Runs 8 and 21, the protoplanets should be significantly smaller than perfectly merged protoplanets in the latter part of the giant impact stage. We will discuss this issue in the next section.

\section{Giant Impact Simulations}\label{sec:GI}

The fragments ejected by a giant impact evolve via successive collisions among them. Small fragments are ground down to the blow-out size, which is $\sim 1~{\rm \mu m}$ around solar-type stars, on a short timescale, whereas large fragments reach the blow-out size later via collisional cascade. Therefore, the decay timescale of the fragments is determined primarily by the mass of the largest fragment and the total ejected mass \citep{Kobayashi2010a}. In this paper, the largest fragment does not refer to the post-impact protoplanet(s), but the largest body ejected by a giant impact. Although \cite{Genda2012} performed giant impact simulations using 20,000 SPH particles, their numerical resolution is insufficient to determine the mass of the largest fragment. Moreover, it is difficult to make a table of the mass of the largest fragment using the data obtained by \cite{Genda2012}, because the formation process of the large fragments is rather stochastic, as shown below. In this section, we perform higher-resolution calculations (100,000 SPH particles) of giant impacts that took place in Run 1, 21, and 26 from the 50 runs, using the exact impact parameters of giant impacts obtained by \cite{Kokubo2010} to precisely calculate the mass of the largest fragment ejected by a giant impact.

To accurately calculate the mass of large fragments, we use the calculation method developed by \cite{Genda2015}. In this method, we use a friends-of-friends algorithm to roughly identify clumps of SPH particles. Then, we iteratively determine if each SPH particle is gravitationally bound to these clumps. Finally we determine if these clumps are gravitationally bound to each other. In this analysis, we set the minimum number of SPH particles for a clump to be 10. If any clumps except for protoplanets are not detected, we set $M_{\rm lrg}$ to be the mass of just one SPH particle.

Our numerical code is the standard SPH method \citep[e.g.,][]{Monaghan1992}, which is the same as that used in \cite{Genda2012}. We performed the SPH simulations of giant impacts between protoplanets with a 30wt\% iron core and a 70wt\% silicate mantle. We applied the Tillotson equation of state \citep{Tillotson1962} in our simulations. The initial spins of the protoplanets were set to zero.

\begin{figure}
\epsscale{1.0}
\plotone{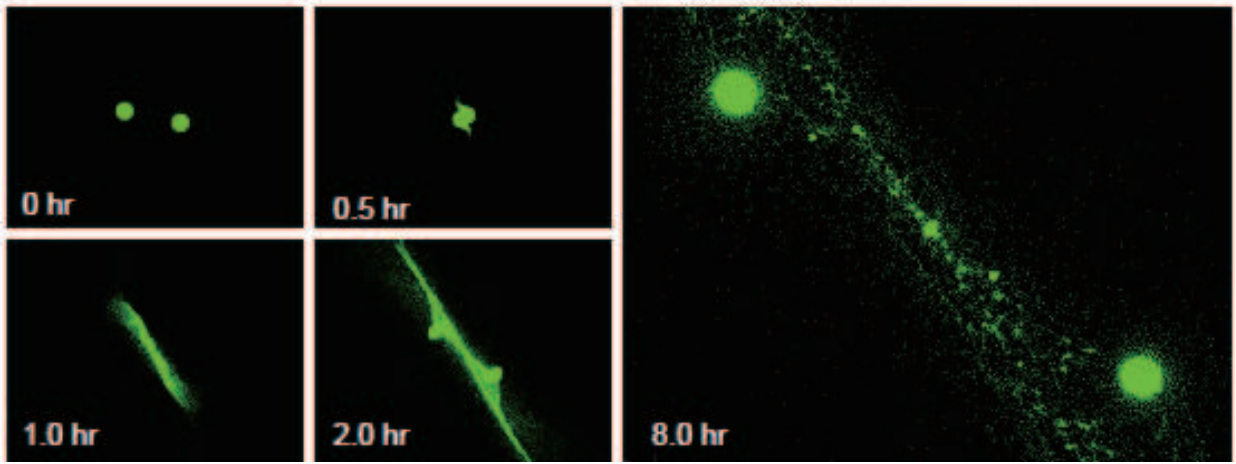}
\caption{Snapshots of a giant impact for the 11th collision in Run 1. This simulation used 100,000 SPH particles. After the first contact, the protoplanets escape from each other and are no longer gravitationally bound. In this collision, a large amount of fragments ($0.05M_{\oplus}$ in total) is ejected. The mass of the largest fragment excluding protoplanets is composed of approximately 2,500 SPH particles, which corresponds to 1/160$M_{\oplus}$ (half of the lunar mass).}
\label{fig:snapshot}
\end{figure}

Figure \ref{fig:snapshot} shows snapshots of the numerical simulation for the 11th giant impact in Run 1 of \cite{Kokubo2010}. After the first contact between two protoplanets, they escape from each other (hit-and-run collision). Numerous SPH particles are ejected by this collision. In this simulation, fragments with a total mass of 0.05$M_{\oplus}$ are ejected. Many clumps are gravitationally formed from ejected SPH particles. The mass of the largest clump (here, we call the largest fragment) is approximately half of the lunar mass, which includes approximately 2,500 SPH particles in this simulation.

\begin{figure}
\epsscale{1.0}
\plotone{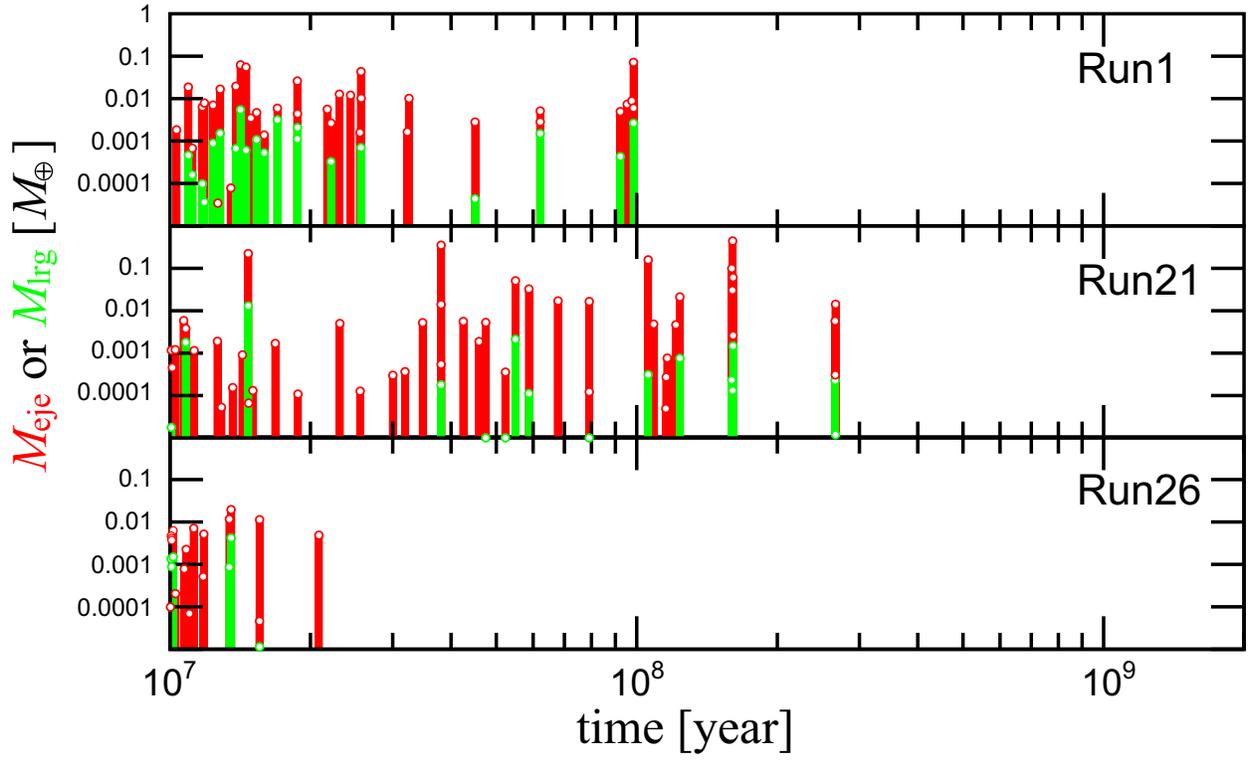}
\caption{Total ejected mass $M_{\rm eje}$ and the mass of largest fragment $M_{\rm lrg}$ produced by each giant impact occuring in Runs 1, 21, and 26.}
\label{fig:Mlrg}
\end{figure}

We performed numerical simulations of all giant impacts that took place in Runs 1, 21, and 26 from the 50 runs, which correspond to the cases of the average ($M_{\rm eje}^{\rm tot} = 0.47M_{\oplus}$), maximum ($M_{\rm eje}^{\rm tot} = 1.76M_{\oplus}$), and minimum ($M_{\rm eje}^{\rm tot} = 0.09M_{\oplus}$) total ejected masses estimated in the previous section, respectively (see Figure \ref{fig:Mejetot}). In Runs 1, 21, and 26 of the simulation, 35, 47, and 17 giant impacts occurred, respectively. Figure \ref{fig:Mlrg} shows a plot of the total ejected mass $M_{\rm eje}$ and the mass of the largest fragment $M_{\rm lrg}$ for each giant impact as a function of time. Calculated $M_{\rm eje}^{\rm tot}$ from these high-resolution giant impact simulations is $0.45M_{\oplus}$, $1.59M_{\oplus}$, and $0.083M_{\oplus}$ for Runs 1, 21, and 26, respectively, which are consistent with previously estimated $M_{\rm eje}^{\rm tot}$ from the table data of low-resolution giant impact simulations. 

As we mentioned in the previous section, the values of $M_{\rm eje}^{\rm tot}$ estimated here is somewhat overestimated. To roughly estimate the effect of the mass loss from protoplanets, we reanalyze $M_{\rm eje}^{\rm tot}$. We use the impact parameters obtained by \cite{Kokubo2010} and collision outcomes obtained from high-resolution impact simulations, but we take into account the effect of the decrease in the mass of protoplanets by giant impacts. The decrease in  $M_{\rm eje}$ is also considered according to the mass of colliding protopalnets whose masses are equal to or smaller than those of perfectly merged protoplanets. We find that the effect of the mass loss reduces $M_{\rm eje}^{\rm tot}$ from $0.45M_{\oplus}$ to $0.38M_{\oplus}$ for Run 1, from $1.59M_{\oplus}$ to $0.82M_{\oplus}$ for Run 21, and from $0.083M_{\oplus}$ to $0.081M_{\oplus}$ for Run 26. The effect of the mass loss is significant for Run 21, and $M_{\rm eje}^{\rm tot}$ is overestimated by a factor of 2. However, significant amount of fragments are still ejected during the giant impact stage. On the other hand, this effect is small for Run 1 and negligible for Run 26.

To compare our results with the stellar age in Section \ref{sec:Discussion}, we assume that the giant impact stage starts at 10 Myr, because the formation of protoplanets followed by formation of planetesimals in a protoplanetary disk would take several Myrs, and the giant impacts are triggered by the gas depletion of protoplanetary disks, which also takes several Myrs inferred from the disk observations \citep{Beckwith1996, Wyatt2003}. As shown in Figure \ref{fig:Mlrg}, most giant impacts occur within 100 Myr. In Run 21, a few disruptive giant impacts ($M_{\rm eje} > 0.1M_{\oplus}$) occur. The mass of the largest fragment $M_{\rm lrg}$ strongly depends on each giant impact event. As shown in Figure \ref{fig:snapshot}, clumping is caused by the mutual gravity of ejected SPH particles. This process is rather stochastic, and it is difficult to estimate $M_{\rm lrg}$ without numerical simulations. As shown in Figure \ref{fig:Mlrg}, some giant impacts result in producing large $M_{\rm lrg}$, and some result in very small $M_{\rm lrg}$. We can safely say that $M_{\rm lrg}$ is less than 10\% of $M_{\rm eje}$ in most cases. In Run 26, almost all giant impacts are gentle events, and after the last giant impact at 20 Myr, the planetary system is stable. 

\section{Lifetime of Debris Disks}\label{sec:Lifetime}

Fragments ejected by a giant impact are further ground down by mutual collisions among them until radiation pressure blows away micron-sized or smaller particles. This collisional cascade reduces the total mass of the fragments $M_{\rm tot}$. The mass evolution of the fragments ejected by a single giant impact via collisional cascade is given by \citep[e.g.,][]{Kobayashi2010a}
\begin{equation}
M_{\rm tot}^{i}(t) = \frac{M_{\rm eje}^{i}}{1+(t-t_{\rm coll}^{i})/t_{\rm dep}^{i}} \;\;\; \;\;\;(t \geq t_{\rm coll}^{i}),
\label{eq:Mtot}
\end{equation}
where $t_{\rm dep}$ is the mass depletion timescale due to collisional cascade and $t_{\rm coll}$ is the time of the giant impact. The superscript {\it i} represents the {\it i}-th giant impact. The size distribution in the steady-state collisional cascade is analytically derived as Eq.~(\ref{eq:q}) in Section~\ref{sec:Excess}, because the mass flux along the mass coordinate is independent of mass \citep{Tanaka1996, Kobayashi2010a}. The mass flux using the steady-state size distribution around $M_{\rm lrg}$ gives $t_{\rm dep}$ as \citep{Kobayashi2010a}, 
\begin{equation}
t_{\rm dep} = 4.2 \times 10^6 
\left( \frac{M_{\rm lrg}}{6 \times 10^{22} {\rm kg}} \right)^{0.64} 
\left( \frac{a}{1 {\rm AU}} \right)^{4.18}
\left( \frac{\Delta a/a}{0.1} \right)
\left( \frac{e}{0.1} \right)^{-1.4}
\left( \frac{M_{\rm eje}}{6 \times 10^{23} {\rm kg}} \right)^{-1} {\rm years},
\label{eq:tdep}
\end{equation}
where $a$ is the semimajor axis at which a giant impact takes place, $\Delta a$ is the range of the semimajor axes of fragments, and $e$ is the typical eccentricity of the fragments. This formulation of the mass depletion timescale is derived by the analysis including erosive collisions caused by low impact energy, which are more effective in the mass evolution due to the collisional cascade, and assuming the collisional strength derived from \cite{Benz1999} for impacts of rocky bodies at a speed of 5 km/s.

The values of $M_{\rm eje}$ and $M_{\rm lrg}$ are derived from the SPH impact simulations in Section \ref{sec:GI}, and $a$ is given by the location of giant impacts obtained by \cite{Kokubo2010}. Because fragments are stirred by the post-collision protoplanet(s), $e$ and $\Delta a$ are estimated from the mass of the protoplanet. We set $e = v_{\rm esc}/v_{\rm K}$ and $\Delta a = 10 r_{\rm H}$, where $v_{\rm esc}$ is the surface escape velocity of the protoplanet ($v_{\rm esc} = \sqrt{2GM_{\rm p}/R_{\rm p}}$, where $M_{\rm p}$ and $R_{\rm p}$ are the mass and radius of the protoplanet, respectively), $v_{\rm K}$ is the Kepler velocity of the protoplanet ($v_{\rm K} = \sqrt{GM_{\rm s}/a}$, where $M_{\rm s}$ is the mass of the central star; here, we use the solar mass), and $r_{\rm H}$ is the Hill radius of the protoplanet ($r_{\rm H} = (M_{\rm p}/3M_{\rm s})^{1/3} a$). If the giant impact is a hit-and-run collision, we use the mass of the larger post-collision protoplanet to estimate $e$ and $\Delta a$.

It should be noted that the stirring by a protoplanet increases the random velocities of fragments up to $v_{\rm esc}$ because fragments with random velocities larger than $v_{\rm esc}$ effectively experience collisions with the protoplanet rather than the stirring. The random velocities, comparable to $v_{\rm esc}$, are much smaller than the Keplerian velocity around 1 AU. Therefore, the scattering ejection of fragments from the central star by protoplanets is negligible. On the other hand, collisions between a protoplanet and fragments reduce the total mass of fragments. The collisional timescale with a protoplanet whose mass is $M_{\rm pro}$ is given by $( n_{\rm s} P_{\rm col} \Omega )^{-1}$, where $n_{\rm s}$ is the surface number density of the protoplanet, $P_{\rm col}$ is the collisional probability, and $\Omega$ is the orbital frequency. Since the number of protoplanets in a fragment annulus is one, $n_{\rm s} \approx 1/2 \pi a \Delta a$. The collisional probability is approximated to be $\pi R^2$ assuming that the relative velocity between fragments and the protoplanet is $v_{\rm esc}$ \citep{Greenzweig1992}. Hence the depletion timescale of fragments by the accretion onto a protoplanet is given by
\begin{equation}
 t_{\rm acc} \approx 1.2 \times 10^{7} 
\left(\frac{M_{\rm pro}}{M_\oplus}\right)^{-2/3} 
\left(\frac{a}{1\,{\rm AU}}\right)^{7/2}
\left(\frac{\Delta a/a}{0.1}\right) \, {\rm years}.
\label{eq:tacc} 
\end{equation}
Because $t_{\rm acc}$ is longer than $t_{\rm dep}$ as shown in Eqs.~(\ref{eq:tdep}) and (\ref{eq:tacc}), the accretion onto protoplanets is negligible. Once the total mass of fragments is smaller than $10^{-2} M_{\oplus}$ due to collisional cascade, the accretion onto a protoplanet mainly reduces fragments. However, the disks with such small fragment masses are much fainter than those with which we are concerned. Therefore, we only consider the depletion of fragments by collisional cascade.

\begin{figure}
\epsscale{1.0}
\plotone{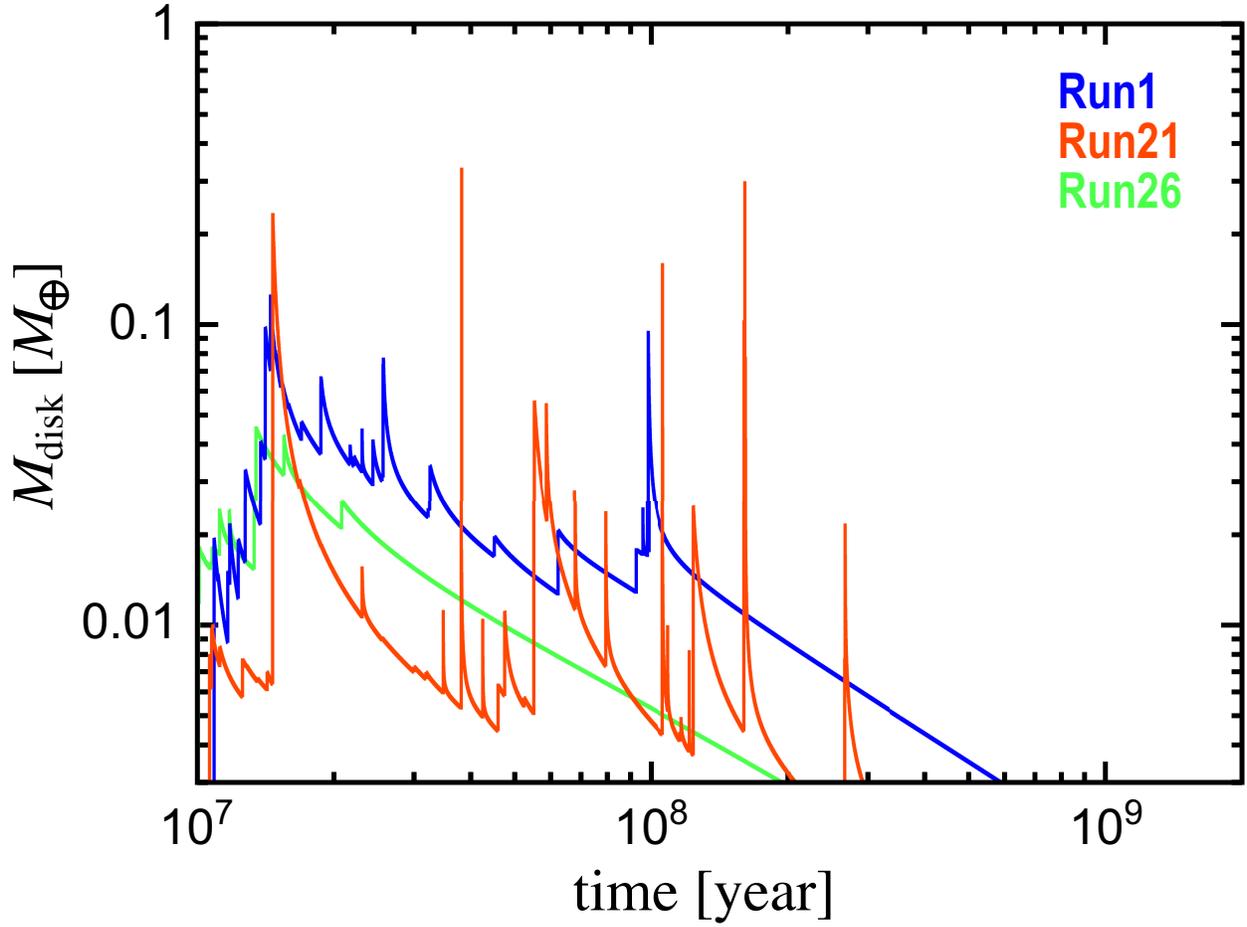}
\caption{Evolution of the mass of the debris disk $M_{\rm disk}$ for Runs 1, 21, and 26. Each spike corresponds to a giant impact, and the decrease in $M_{\rm disk}$ is caused by collisional cascade.}
\label{fig:Mdisk}
\end{figure}

A collision between protoplanets ejects fragments, and the total mass of the fragments decreases by collisional cascade. Another subsequent impact increases the total mass of fragments. Because each $t_{\rm dep}$ is generally shorter than the time intervals of the giant impacts, for simplicity, we deal independently with the mass evolution of fragments produced by each giant impact. The total mass of the debris disk $M_{\rm disk}$ as a function of time throughout the giant impact stage is given by the summation of $M_{\rm tot}^{i} (t)$ over all giant impact events,
\begin{equation}
M_{\rm disk}(t) = \sum_{i} M_{\rm tot}^{i}(t).
\label{eq:Mdisk}
\end{equation}
This independent treatment makes $t_{\rm dep}$ slightly overestimated, because overlapped region of debris disks produced by giant impacts has larger $M_{\rm eje}$. Figure \ref{fig:Mdisk} shows the evolution of $M_{\rm disk}$ for three runs. Here, we assume that the giant impact stage begins at 10 Myr. Each spiky structure in the figure corresponds to a giant impact event. The collisional cascade following a giant impact reduces $M_{\rm disk}$, whereas subsequent giant impacts increase $M_{\rm disk}$. As a result, repetitive giant impacts ensure $M_{\rm disk}$ remains larger than 0.01$M_{\oplus}$ for Runs 1 and 26, and larger than 0.003 $M_{\oplus}$ for Run 21 until $\sim$ 100 Myr.

\section{Infrared Excess of Debris Disks During Giant Impact Stage}\label{sec:Excess}

In collisional cascade, the size distribution of the fragments is determined by collisional equilibrium. The differential surface number density of fragments with mass $m$, defined as $n_{\rm s}(m){\rm d}m$, is proportional to $m^{-q} {\rm d}m$, and the index $q$ is given by \citep{Kobayashi2010a}
\begin{equation}
q = - \frac{11+3p}{6+3p},
\label{eq:q}
\end{equation}
where $p = {\rm d} [\ln{(Q_{\rm D}^{*}/v^2)}]/{\rm d} [\ln{m}]$, $Q_{\rm D}^{*}$ is the specific impact energy needed for the ejection of half mass of the colliders, and $v$ is the impact velocity between bodies. More massive bodies tend to have larger $Q_{\rm D}^{*}$ values in the gravity regime (for kilometer-sized or larger bodies) because of the strong gravity, whereas $Q_{\rm D}^{*}$ is constant or has a negative dependence on $m$ in the strength regime (for kilometer-sized or smaller bodies). Because $v$ is stirred by protoplanets, $v$ is independent of $m$. Therefore, using the mass dependence of $Q_{\rm D}^{*}$ \citep{Benz1999}, we set $q = 1.68$ for $s > s_0$ and $q = 1.89$ for $s < s_0$, where $s$ is the radius of the body and $s_0$ = 1 km is the approximate value determined from the simulation by Benz and Asphaug (1999). In addition, the micron-sized or smaller grains in the debris disks around the solar-type stars are blown out by radiation pressure on a short timescale comparable to the Keplerian period ($\sim$ 1 year), and hence we set the smallest size of this size distribution as 1 $\mu$m. Using this size distribution, we obtain the vertical optical depth of the debris disks $\tau$ from $M_{\rm disk}$. 

For the warm debris disks considered in this study, the peak wavelength of thermal emission is not much longer than the smallest dust in the debris disks, which is determined by radiation pressure blow-out. We thus approximate the thermal emission of dust as blackbody radiation. In addition, debris disks are optically thin, because $\tau$ just after a giant impact is typically $<$ 0.1 at the orbit where a giant impact takes place, and $\tau$ decreases quickly with time. Thus, we can express the emission flux with the wavelength $\lambda$ from the debris disk as
\begin{equation}
F_{\nu}^{\rm disk}(\lambda) = \frac{1}{D^2} \int 2 \pi r \tau (r) B_{\nu} (T) {\rm d}r,
\label{eq:Fdisk}
\end{equation}
where $D$ is the distance of the star from the Earth, $r$ is the radial distance of the debris disk from the central star, and $B_{\nu}$ is the Planck function for a given temperature $T$. The temperature is defined as $T = 280 (r/1{\rm AU})^{1/2}$ K. Equation (\ref{eq:Fdisk}) is integrated over the debris disk, typically from 0.3 AU to 2 AU.

The summation of the disk flux $F_{\nu}^{\rm disk} ({\lambda})$ and the stellar flux $F_{\nu}^{*} ({\lambda})$ gives the observational flux $F_{\nu}^{\rm obs} ({\lambda})$. Figure \ref{fig:F24} shows the flux ratio $F_{\nu}^{\rm obs} (24 \mu {\rm m})/F_{\nu}^{*} (24 \mu {\rm m})$ for Runs 1, 21 and 26, where we use the solar value of $F_{\nu}^{*} (24 \mu {\rm m})$. A giant impact increases the flux ratio, which is decreased by the subsequent collisional cascade. Multiple impacts lead to the spiky evolution of the flux ratio, which is similar to the mass evolution of the debris disk shown in Figure \ref{fig:Mdisk}. The calculated observational flux is comparable to the stellar flux until 100 Myr. Our results indicate that bright warm debris disks are generally formed if planetary systems have a giant impact stage, as our solar system had.

\section{Discussion and Conclusions}\label{sec:Discussion}

\begin{figure}
\epsscale{1.0}
\plotone{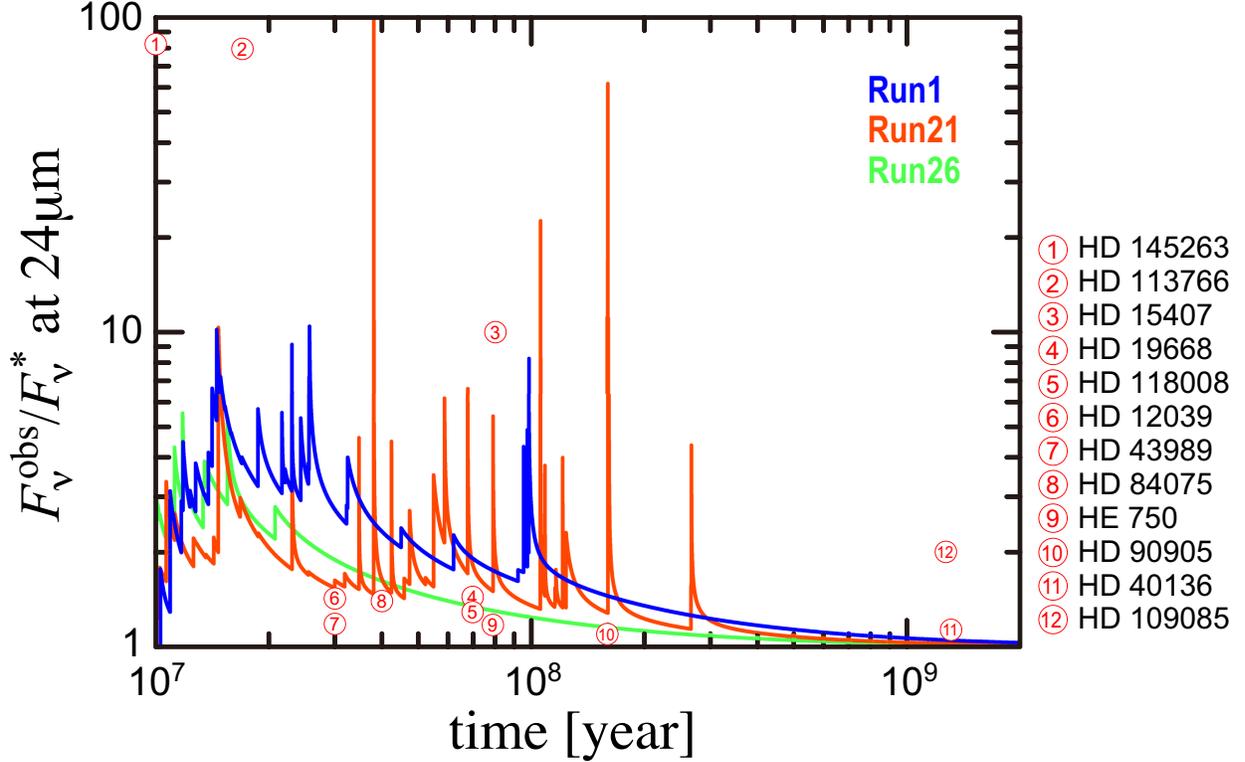}
\caption{Calculated observable 24 $\mu$m flux scaled by the stellar 24 $\mu$m flux as a function of time. The observable flux $F^{\rm obs}_{\nu}$ is the summation of the debris disk's flux $F^{\rm disk}_{\nu}$ and the stellar flux $F^{\rm *}_{\nu}$. Observational data of infrared excess are from \cite{Chen2011} for stars labeled with numbers 1 and 2, \cite{Melis2010} for number 3, \cite{Zuckerman2011} for numbers 4 to 8, \cite{Carpenter2009} for numbers 9 and 10, and \cite{Beichman2006} for numbers 11 and 12.}
\label{fig:F24}
\end{figure}

In the late stage of terrestrial planet formation in our solar system, Mars-sized protoplanets that formed through the accretion of planetesimals collide with each other because of the chaotic orbital instability after gas depletion, resulting in the formation of Earth and Venus. This scenario is consistent with the core formation timescale of Earth ($\sim$ 100 Myr) and Mars ($<$ 15 Myr) obtained from Hf/W isotope analysis \citep[e.g.,][]{Halliday2008, Dauphas2011, Kobayashi2013}. The final giant impact in the formation of Earth may result in the formation of the Moon \citep[e.g.,][]{Canup2004}. We estimated the total mass of the fragments ejected by giant impacts using the numerical results of \cite{Kokubo2010} and \cite{Genda2012}. We found that a significant amount of fragments ($\sim 0.4 M_{\oplus}$ in total) is ejected into the terrestrial planet region throughout the giant impact stage ($\sim$ 100 Myr). Moreover, we accurately calculated the mass of the largest fragment using high-resolution SPH simulations of giant impacts. Using these results, we obtained the evolution of the debris disks throughout the giant impact stage. A single giant impact produces some amount of fragments, and their total mass is decreased by collisional cascade among these fragments. A series of giant impacts forms debris disks, and the disk's infrared flux has a spiky time evolution from 10 Myr to at least 100 Myr (see Figure \ref{fig:F24}).

Our simulation results can be compared with observational results. The observed 24 $\mu$m flux ratios for warm debris disks ($>$ 150 K) are also plotted in Figure \ref{fig:F24}. The debris disks caused by giant impacts can explain most of the observed warm debris disks around stars with ages of 10 -- 100 Myr. However, HD 145263 and HD 113766, which have ages of $\sim$ 10 Myr, have extremely bright disks with flux ratios of $\sim$ 100, and our calculated evolution of infrared excesses during the giant impact stage cannot readily explain their existence. There is a possibility that disruptive giant impacts that occurred at 40 Myr and 150 Myr in Run 21 took place just at the ages of HD 145263 and HD 113766. In addition, if vapor condensates produced by high-velocity collisions between protoplanets are considered, extremely sharp spikes in infrared excess would be expected. This is because vapor materials re-condense into small droplets with characteristic sizes of a few mm to a few cm \citep[e.g.,][]{Melosh1991, Johnson2012, Johnson2014}, and these mm-cm size spherules grind away very quickly and produce huge infrared excess. However, because an excess duration of flux ratios of 100 is very short ($<$ 1000 yrs), the observed probability should be extremely low. On the other hand, HD 109085, which is $\sim$ 1 Gyr in age, has a debris disk with a flux ratio of $\sim$ 2, which also cannot be explained by the giant impact stage, because the final impact usually occurred earlier than 300 Myr in the formation of our terrestrial planets. From a theoretical point of view, it is difficult to prolong the giant impact stage up to 1 Gyr in our solar system. Another mechanism may be required to describe the  debris disk formation after terrestrial planet formation, such as the Nice model \citep{Gomes2005}, in which gas giant planets move after planet formation.

Debris disks are also formed during the growth of protoplanets through the accretion of planetesimals because the stirring by protoplanets leads to destructive collisions between planetesimals \citep{Kenyon2002, Kennedy2010, Kobayashi2014}. In systems where the giant impact stage is expected, the collisional destruction of planetesimals during protoplanet growth forms cold debris disks around stars with ages of $\sim$ 10 -- 100 Myr. We can distinguish between planetesimal-induced cold and giant impact-induced warm debris disks. However, if planetesimal disks are less massive, the formation and growth of protoplanets in the terrestrial planet formation region occur even at 10 -- 100 Myr, resulting in warm debris disks around stars with ages of 10 -- 100 Myr \citep{Kobayashi2014}. The time evolution of disk fluxes originating from the formation and growth of protoplanets is smooth, which is quite different from the debris disks caused by giant impacts. We will be able to determine major origin of warm debris disks when the number of sampled debris disks is sufficiently high. Spectral feature of warm debris disks produced by giant impacts would be also different from that produced by planetesimal collisions \citep[e.g.,][]{Lisse2009}. This is because protoplanets should be differentiated, and fragments ejected by giant impacts would be rich in crustal component, which is different from the composition of undifferentiated planetesimals.

Our results show that debris disks produced by giant impacts can be observable at 24 $\mu$m infrared excess throughout the giant impact stage ($\sim$ 100 Myr). However, not all stars with ages of 10 -- 100 Myr have infrared excess. This observational result might suggest that the giant impact stage that we consider in this paper is not very common in extrasolar systems. Observations of FGK stars with ages of 10 -- 100 Myr by the {\it Spitzer} space telescope have shown that the fraction of stars with 24 $\mu$m infrared excess ranges from 8\% \citep{Beichman2006, Carpenter2009, Trilling2008} to 48\% \citep{Zuckerman2011}. Observation of these stars with multiple wavelengths gives the temperature of the debris disk. The fraction of stars with the component of a warm debris disk ($>$ 150K) among the stars with 24 $\mu$m infrared excess ranges from 13\% \citep{Jackson2012} to 36\% \citep{Morales2011}. Therefore, the fraction of stars with a warm debris disk is 1 -- 17\% in total, which is also consistent with the recent observations at 12 $\mu$m by {\it WISE} space telescope \citep{Kennedy2013}. As described above, the giant impact stage in the solar system is supported by theoretical studies \citep[e.g.,][]{Wetherill1985, Kokubo1998, Jacobson2014} and isotope chronometry \citep[e.g.,][]{Touboul2007, Halliday2008}. Therefore, these observations indicate that the giant impact stage during which a few Earth-sized planets form around 1 AU is not very common around other stars, probably $\sim$ 10\%, which is consistent with the estimate made by \cite{Jackson2012}.

Because many extrasolar planets have been discovered so far, we can compare our results with the planet occurrence rates of extrasolar systems. In particular, thanks to the {\it Kepler} space telescope, the occurrence rates of Earth-sized planets and super-Earths have been recently investigated. Although there is not sufficient data about Earth-sized planets and super-Earths located around 1 AU, recent {\it Kepler} data show that the occurrence rate of planets with radii of $1 R_{\oplus}  < R < 2 R_{\oplus}$ and with orbital periods of $150 < P < 250$ days is $9.6 \pm 3.3 \%$ for MKGF stars \citep{Mulders2015}, where $R_{\oplus}$ is the radius of the Earth. This reported rate seems to be consistent with our speculation that the giant impact stage as experienced in our solar system is not very common.

Because many short-period super-Earths have been discovered, the occurrence rates of these planets are reliable. By using {\it Kepler} data, \cite{Petigura2013} found that the occurrence rate of planets with radii of $1 R_{\oplus}  < R < 2 R_{\oplus}$ and with orbital periods of $5 < P < 100$ days is $26 \pm 3 \%$. Here, we consider the case of giant impacts that make super-Earths located at less than 0.4 AU (which corresponds to an orbital period of 100 days around a G-type star). Because the orbital period at 0.4 AU is shorter than that at 1 AU, the orbital instability among protoplanets that causes giant impacts takes place very quickly. Therefore, the timescale of the giant impact stage for these planetary systems is much shorter than that of our terrestrial planets. Moreover, to make super-Earths in situ, the protoplanets should be larger than Mars-sized protoplanets \citep{Kokubo2002}. Larger protoplanets also shorten the timescale of the giant impact stage \citep{Kokubo2006}. Therefore, even if giant impacts take place in the systems of short-period super-Earths, the duration of the giant impact stage should be much shorter ($\sim$ 1 Myr) than that for our terrestrial planets ($\sim$ 100 Myr). Moreover, the lifetime of debris disks produced by giant impacts should be also very short, because it strongly depends on the distance of the debris disk from the central star (see Eq. (\ref{eq:tdep})), although the effect of large $M_{\rm lrg}$, $\Delta a$ and $e$ on prolonging the lifetime of debris disks should be considered. The lifetime of a debris disk around 0.4 AU is 50 times shorter than that around 1 AU. Therefore, the probability of observing a giant impact stage for short-period super-Earths is extremely low. One additional point that might support non-detectability of super-Earth giant impacts is that collisions between super-Earths likely produce more vaporized ejected materials, because of their larger escape velocities and higher impact velocity. Because the size of condensed materials would be very small, the lifetime of the debris disks produced by collisions between super-Earths would be shorter than that produced by collisions between Earth-sized objects. Therefore, it is likely that observed warm debris disks are the result of the giant impact stage of our solar system-like terrestrial planet formation.

\acknowledgments

HG and HK gratefully acknowledge support from Grant-in-Aid for Scientific Research (B) (26287101). This work was also supported by Research Grant 2015 of Kurita Water and Environment Foundation. Numerical simulations were in part carried out on computers at the Center for Computational Astrophysics, National Astronomical Observatory of Japan.

\clearpage



\clearpage

\clearpage

\clearpage

\clearpage

\clearpage

\clearpage









\clearpage

\end{document}